\documentstyle[prb,aps,epsfig]{revtex}

\begin{document}

\draft

\title{ Magnons in CMR pyrochlore Tl$_2$Mn$_2$O$_7$}

\author{C. I. Ventura}

\address{Centro At\'omico Bariloche, 8400 - Bariloche, Argentina}

\author{M. Acquarone}

\address{IMEM-CNR, INFM - Unit\'a di Parma, Dipartimento di Fisica, 
Universit\`{a} di Parma, Italy.}

\maketitle

\begin{abstract}
Well defined spin waves were observed when the spin dynamics 
of Tl$_2$Mn$_2$O$_7$, the first pyrochlore compound 
found to exhibit colossal magnetoresistance, was measured  
[J.W.Lynn et al., 
Phys. Rev. Lett. {\bf 80},4582(1998)], 
in stark contrast with the experimental results on the larger 
family of magnetoresistive manganites with perovskite structure. 
In this work, we present our calculation for the spin waves 
in Tl$_2$Mn$_2$O$_7$, which we described using the microscopic 
generic model proposed recently for this compound 
[C.I.Ventura and M.A.Gusmao, Phys. Rev. B {\bf 65}, 14422(2002)]. 
We have employed a canonical transformation to determine
perturbatively the effective spin-wave Hamiltonian, 
obtaining therefrom the renormalization of the ferromagnetic spin
waves related to the localized Mn$^{4+}$ spins, due to their coupling    
with the conduction electrons present. We have calculated the
magnon dispersion relations along different paths in the 
first Brillouin zone, comparing them with those which are 
obtained for an ideal isotropic ferromagnet. This comparison 
evidences an agreement between the ferromagnetic magnons obtained 
from the generic model and the bare spin waves, such as had been found  
in neutron scattering experiments.    
    
\end{abstract}





\section{INTRODUCTION}
\label{sec:intro}

During the eight years passed since the first discovery of colossal magnetoresistance (CMR) in 
 a manganite compound with pyrochlore crystal structure,  
Tl$_2$Mn$_2$O$_7$,\cite{5,6,7} increasing evidence has been found of the many differences 
in physical properties with respect to the much better known family of CMR
 Mn-perovskites.\cite{3}  
Starting from the different crystal structures, which in the
 pyrochlore case includes a tetrahedral 
(instead of cubic) array of the MnO$_6$ octahedra, with the possibility of 
the presence of magnetic frustration effects. Also the larger bend in the Mn-O-Mn bond angle 
(133 vs 170 degrees in perovskites) would tend to reduce the magnitude of any antiferromagnetic 
(AF) superexchange coupling.\cite{7} The lack of Jahn-Teller distortions in the pyrochlores rules out
the presence of Mn$^{3+}$,\cite{5,7,single} and therefore of the
 double-exchange mechanism proposed in connection with the perovskite Mn oxides.     
Another difference lies in the small spin-lattice correlation found in pyrochlores, 
where no abrupt changes in the lattice parameters appear near the
 critical temperature (Tc$\sim 125 K$,\cite{5,6,7}) 
being the  MnO$_6$ octahedra almost temperature-independent while only varying smoothly their tilting angles
inside the structure.\cite{5} A large difference has been found early on, regarding the numbers of carriers:
while CMR in perovskites is found to be optimal for about 30 percent doping, Hall experiments 
in Tl$_2$Mn$_2$O$_7$ \cite{7,sushkolast} indicated only about 10$^{-3}$ to 10$^{-5}$ conduction 
electrons per formula unit (with a one third reduction above Tc). No anomalous Hall contribution 
is present in the pyrochlores. We can also mention the opposite behaviour of Tc upon the 
application of pressure:\cite{presion,boston} 
decreasing in the pyrochlores (even with a reversal of tendency at high pressures). 

The large amount of experimental evidence gathered, including also studies of compounds where 
the different components of  Tl$_2$Mn$_2$O$_7$ 
have been doped by
substitution,\cite{6,rudoping,scdoping,sbdoping,bidoping,Odefect,cddoping} 
put into serious 
question if a common underlying mechanism for CMR could be present for perovskites and the pyrochlores,
where transport and magnetism seem to be primarily related to
different electronic orbitals,\cite{7,scdoping}  
coupled by hybridization. 
A series of theoretical proposals have been put forward and explored in connection with these results 
in pyrochlores.\cite{civba,littlewood,dolores,civmag} In particular, 
the first microscopic model studied for Tl$_2$Mn$_2$O$_7$ was 
an intermediate valence model (IVM)  \cite{civba} based on the suggestion \cite{7} of the possibility of 
the presence of a small (x$\sim$0.01, related to the small number of
carriers) amount of internal doping of the 
type  Tl$^{3+}_{2-x}$Tl$^{2+}_{x}$Mn$^{4+}_{2-x}$Mn$^{5+}_{x}$O$_7$. The IVM \cite{civba} proved useful 
for the description of the main magnetotransport measured characteristics and the initially puzzling 
Hall data. This 
was done through the temperature- and magnetic field-dependent changes of the electronic structure,
which in the ferromagnetic phase exhibited common features with the available band structure 
calculations,\cite{bsub,bsingh} including hybridization gaps in all bands except for the 
minority-spin carrier band. In the paramagnetic phase, all bands had developed spin-independent gaps 
and the reduction in the number of effective carriers could be explained placing the Fermi level
slightly above the hybridization gap.\cite{civba}           
Later, the generic model for Tl$_2$Mn$_2$O$_7$ \cite{civmag} was put forward to explore a whole set of 
other proposals suggested for this
compound.\cite{5,7,presion,littlewood,mishra,dolores} 
This model includes a band of itinerant carriers
hybridized with other strongly correlated electronic orbitals. These appear forming a lattice of  
 Mn$^{4+}$ local magnetic moments, as well as occupying a narrow band, 
coupled through a strong Hund interaction.\cite{civmag} A superexchange interaction between the 
Mn$^{4+}$ spins is also included. The electronic structure was shown to 
share the main features of that obtained previously with the IVM
model,\cite{civba}  when using appropriate sets of
parameters.\cite{civmag}  

In the present work, we address the problem of the study of the spin excitations in Tl$_2$Mn$_2$O$_7$,
which were also found to be in stark contrast to those observed in perovskite manganites when 
 measured in neutron scattering experiments.\cite{lynn} The excitations observed in Tl$_2$Mn$_2$O$_7$  
are magnons characteristic of a soft three-dimensional isotropic ferromagnet, softening at higher 
temperatures and appearing to collapse as the critical temperature is approached, being the 
ferromagnetic transition driven by thermal  population of conventional spin waves.    
Here, we have calculated the spin wave dispersion relations for Tl$_2$Mn$_2$O$_7$ along different 
trajectories in the first Brillouin zone (BZ), using the 
generic model \cite{civmag} with parameters found useful before for the description of the compound, 
and assuming a ferromagnetic superexchange interaction to be present. 
In Section II an outline  is given of the method we have used to calculate the magnons: 
basically, we employed a canonical transformation to determine perturbatively 
the renormalization of the ferromagnetic spin waves related to the Mn$^{4+}$ local moments, 
by their Hund coupling with the conduction electrons.            
In Section III, the general analytical expressions obtained in the previous section 
are evaluated for the specific case of Tl$_2$Mn$_2$O$_7$. A complete presentation 
and discussion of our results is included. 
A comparison of the renormalized spin waves calculated   with the bare magnons, 
characteristic of a three-dimensional isotropic ferromagnet 
and alike to those measured in the available neutron scattering
experiments,\cite{lynn} 
evidences a good agreement. In Section IV a summary of our study and its conclusions 
are included, together with suggestions for new experiments which could test 
some of our predictions for other yet unexplored cases.

\section{EFFECTIVE SPIN-WAVE HAMILTONIAN FOR  T\lowercase{l}$_{2}$M\lowercase{n}$_{2}$O$_{7}$}
\label{sec:spinwaves}

We will begin by outlining our general scheme of calculation. 
The Hamiltonian as usual is split into  non-interacting ($H_0$) and
interacting ($V$) terms: 
\begin{equation} 
 H \, = \, H_{0} \, + \, V \, 
\end{equation}
Next we apply a unitary transformation yielding:   
\begin{equation}
\tilde{H} \, = \, e^{i S} H e^{-i S} \, , 
\qquad S=S^{\dagger} 
\end{equation}
Developing $\tilde{H}$ to second order of
perturbations we arrive at the following effective Hamiltonian:   
\begin{equation}
 \tilde{H} \,\, \sim \,\, \tilde{H}^{(2)} \,\, =  \, \, H_{0} \, \, 
+ \, \, \frac{i}{2} \,  [ S, V ] \,\, + \, 0(\,S^{3}\,) \,\, ,   
\label{h2}
\end{equation}
where  $ S$ obeys: 
$  V \,\, + \,\, i \,\, [ S , H_{0} ] \,\, = \,\, O \,\,. $
Integrating this constraint in the interaction representation 
 $ ( {\cal{O}_{I}}(t) \, \equiv \, e^{i H_{0} t } {\cal{O}} e^{-i H_{0} t}
\, )$, yields:
 \begin{equation}
 S \,\, \sim \,\, S_{I}(t=0) \,\, =  \, \,  \, 
 \frac{1}{\hbar} \,\, \int_{-\infty}^{0} \, \, dt \, \,  V_{I}(t) \,\,
 .
\label{si}
\end{equation}

For the present problem, we consider  the generic
 model recently proposed \cite{civmag} to describe Tl$_2$Mn$_2$O$_7$,
which includes Hund and superexchange couplings as well as 
hybridization effects. Two kinds of electronic orbitals are present, 
as widely believed to be the case: 
one, directly related to the magnetism of the compound, 
involving localized Mn$^{4+}$ moments and a
narrow band strongly Hund-coupled to them.\cite{civmag}  
The other, corresponding to more extended
electronic orbitals, is related to the carriers which hybridize with the narrow band.
A superexchange coupling between the localized
spins is also present. The Hamiltonian \cite{civmag} reads:
 \begin{equation}
H \,   =   \, H_{c} \, + \, H_{d} \, + \, H_{cd}  \, + \, H_{S}
\, , \, 
\label{Hgeneric}
\end{equation}
where:
\begin{eqnarray}
H_c  &=&  - t_{c} \sum_{\langle i,j\rangle \sigma}   
(\, c^\dagger_{i\sigma} c^{}_{j\sigma}  \, + \,  \textrm{H.c.} \,)  - \, \mu   
\, \sum_{i\sigma} \, c^\dagger_{i\sigma} c^{}_{i\sigma} \; , \nonumber \\
H_d  &=&  - t_{d} \sum_{\langle i,j\rangle \sigma}   
(d^\dagger_{i\sigma} d^{}_{j\sigma}  +  \textrm{H.c.})  +
(\epsilon_d - \mu)    
\sum_{i\sigma} d^\dagger_{i\sigma} d^{}_{i\sigma}  \; , \nonumber \\
H_{cd}  &=&  t_{cd} \sum_{i\sigma} \left(   
 c^\dagger_{i \sigma} d^{}_{i\sigma} + d^\dagger_{i\sigma} c^{}_{i\sigma}
  \right) \; , \nonumber \\
H_S  &=&  \, -   J_{\rm H}  \sum_i {\bf S}_i \cdot
{\bf s}^{(d)}_i - J_S \sum_{\langle i,j\rangle} {\bf S}_i \cdot
{\bf S}_j \; . 
\label{Hparts} 
\end{eqnarray}
Here $H_{c}$ describes the band associated \cite{civmag} to the 
weakly  correlated carriers present in Tl$_{2}$Mn$_{2}$O$_{7}$, 
characterized by a hopping parameter $ t_{c}$ between nearest
neighbors, and creation operators
$c^\dagger_{i\sigma}$ for electrons with spin $\sigma$ in the
respective Wannier orbital centered on site $i$.  $H_{d}$ describes
the narrow band associated to the strongly correlated
orbitals,\cite{civmag} with hopping  $ t_{d} (<t_{c}) $, 
and creation operators $d^{\dagger}_{i\sigma}$ 
for electrons with spin $\sigma$ in the
Wannier orbital of energy $\epsilon_{d}$ (with respect to
the center of the $c$ band,$\epsilon_{c} = 0$) localized on site $i$. The chemical
potential of the system is denoted by $\mu$, and is determined by the
total electronic filling $n$. The hybridization  $H_{cd}$
 between $c$ and $d$ electron bands is assumed to be local.
 $H_{S}$ denotes the spin Hamiltonian, which includes the Hund (or
Kondo-like) coupling, measured by $J_{H}$, between a local magnetic moment $ {\bf
{S}}_{i}$, of magnitude
$M = 3/2 $ associated to a Mn$^{4+}$ ion at site $i$,    
and the itinerant spin $s=1/2$
denoted by ${\bf {s}}^{(d)}_{i}$, corresponding to the $d$ 
electron orbitals related to Mn (hybridized with O) in
Tl$_{2}$Mn$_{2}$O$_{7}$.\cite{civmag} A Heisenberg exchange term is also included
in the spin Hamiltonian to allow for the superexchange  
coupling, measured by $J_{S}$, effective between the local moments of
nearest neighbor Mn$^{4+}$ ions.
 
Given this Hamiltonian (\ref{Hgeneric}) we identify $H_0$  as follows:
 
  \begin{eqnarray}
 H_{0} \,& = & \, H_{0}^{e} \, + \, H_{0}^{b} \, \, \nonumber \\ 
  H_{0}^{e} \,&  = & \, \sum_{q,\sigma} \,  \,\left\{ \epsilon_{c}(q) \, 
c^{\dagger}_{q \sigma} c_{q \sigma} \, + \, 
[ \epsilon_{d}(q) - \frac{\sigma J_{H} M}{2} ] \, 
d^{\dagger}_{q \sigma} d_{q \sigma} \, + \, t_{cd} \, ( \, 
c^{\dagger}_{q \sigma} d_{q \sigma} \,+ \,
 d^{\dagger}_{q \sigma} c_{q \sigma} \,)  \,\right\} \nonumber \\
 &  \equiv & \, \sum_{q,\sigma} \, [ \, E^{\alpha}_{q \sigma} \, 
\alpha^{\dagger}_{q \sigma} \alpha_{q \sigma} \, + \, 
E^{\beta}_{q \sigma} \, \beta^{\dagger}_{q \sigma} \beta_{q \sigma} \, ] \, 
\, \nonumber \\
\end{eqnarray}

where  the bare tight binding energies read  
$ \, \epsilon_{\gamma}(q) =
\epsilon_{\gamma} - t_{\gamma} \sum_{n=1}^{z} e^{-i {\bf q}\cdot {\bf a}_{n}}$ 
( with $\gamma =c,d$ ,  and ${\bf a}_{n}$ 
denoting the vectors connecting a site with its $z$ nearest neighbors), while we 
have introduced the electronic eigenbands: 
\begin{eqnarray}
  E^{\stackrel{\alpha}{\beta}}_{q \sigma} \,& = & \,\frac{ \epsilon_{c}(q) \, 
+ \, \tilde{\epsilon}_{d}^{\sigma}(q) \stackrel{-}{+} \, \sqrt{  
( \epsilon_{c}(q) \, - \, \tilde{\epsilon}_{d}^{\sigma}(q) )^2 + 4 t_{cd}^2 } }{2} \, , 
\nonumber \\
 \tilde{\epsilon}_{d}^{\sigma}(q) \,& = & \,[ \epsilon_{d}(q) -   
\frac{\sigma J_{H} M}{2} ] \, , \, 
 \end{eqnarray}
\begin{equation}
\left( 
\begin{array}{c}
d^{\dagger}_{q \sigma} \\
c^{\dagger}_{q \sigma} 
\end{array}
\right) 
\,  =  \, 
\left( 
\begin{array}{cc}
\cos \eta_{q \sigma} & \sin \eta_{q \sigma}  \\ 
- \sin \eta_{q \sigma}  & \cos \eta_{q \sigma}
\end{array}
\right) 
\left( 
\begin{array}{c}
\alpha^{\dagger}_{q \sigma} \\
\beta^{\dagger}_{q \sigma}
\end{array}
\right) \,  , 
\end{equation}


with: 
$$ \tan (2  \eta_{q \sigma}) \, = \, \frac{2 t_{cd}}{
 \epsilon_{c}(q) \, - \, \tilde{\epsilon}_{d}^{\sigma}(q)} \, . $$

We have assumed a ferromagnetic superexchange interaction, consistently 
with the experimental neutron scattering data \cite{lynn} indicating spin excitations 
characteristic of an isotropic ferromagnet.  
Using the Holstein-Primakoff transformation  the 
local moments  are expressed in term of bosonic ferromagnetic  
magnon operators as: $
{\bf S}_{l}^{{}}=\left( S_{l}^{+},S_{l}^{-},S_{l}^{z}\right) =
\left( \sqrt{2M
}b_{l}^{{}},\sqrt{2M}b_{l}^{\dagger },M-b_{l}^{\dagger
}b_{l}^{{}}\right) \, $, at low temperatures.   
The superexchange term in $H_{S}$  is then rewritten  as: 
\begin{eqnarray}   
 H_{0}^{b} \, & = & \, - \frac{J_{S} M^2 z N}{2} \, + \, \sum_{q} \, 
\, \omega_{q} \,\,  b^{\dagger}_{q}\, b_{q}  \, , 
\end{eqnarray}
where $\omega_{q}$ is the bare ferromagnetic magnon energy 
on a lattice of $N$ sites.

On the other hand, $V$ is then identified as the Hund term of the
Hamiltonian, that is:
\begin{eqnarray}
V \, & = & \, \, -J_{H} \, \, \sum_{l} \, \, \left[  
s_{l}^{z}S_{l}^{z} \, + \, \frac{1}{2}
\left( s_{l}^{+}S_{l}^{-}+s_{l}^{-}S_{l}^{+}\right) \, \right] \,
\equiv \, V_{z} \, + \, V_{xy} \, , 
\end{eqnarray}
where the longitudinal ( $V_z$ ) and transversal ( $V_{xy}$ ) terms  
respectively are:
\begin{eqnarray}
V_{z} \,  & = & \, 
 - \frac{J_{H} M}{2}\, \sum_{k} \, \left(
n_{k\uparrow }^{d}-n_{k\downarrow }^{d}\right) \,  
+ \, \frac{J_{H} }{2N} \, \sum_{kpq, \sigma}  \, \, \sigma \, \, 
\,  d_{k\sigma }^{\dagger }d_{p+q\sigma}^{{}} 
b_{q}^{\dagger } b_{k-p}^{{}}
 \, , \\
V_{xy} \, & = & \, - 
  J_{H} \, \sqrt{\frac{M}{2 N}} \, \sum_{kq} \, \left( d_{k\uparrow }^{\dagger
}d_{k+q\downarrow }^{{}}b_{q}^{\dagger } 
+d_{k\downarrow }^{\dagger}d_{k+q\uparrow }^{{}}b_{-q}^{{}}
 \, \right) \, . 
\end{eqnarray}
Here we employ the usual notation for Fermi number operators: $n^{\gamma}_{k\sigma} \, = \, 
\gamma^{\dagger}_{k,\sigma} \gamma_{k,\sigma}$, in this case with $\gamma = d$.

To avoid divergences in Eq.~(\ref{si}), we have redefined $H_{0}^{e}$
by incorporating the first term of  $V_{z}$ as a correction to the 
$d-$band energies, and continued working with the following effective 
interaction:
\begin{eqnarray} 
 V'\, & = & \, V'_{z} + V_{xy} \, , \\
V'_{z} \, & = & \, \, \frac{J_{H} }{2N} \, \sum_{kpq, \sigma}  \, \, \sigma \, \, 
\,  d_{k\sigma }^{\dagger }d_{p+q\sigma}^{{}} 
b_{q}^{\dagger } b_{k-p}^{{}}
\, .
\label{veff}
\end{eqnarray}

By using $V'$ in  Eq.~(\ref{si}), we obtain:
\begin{eqnarray}
 S \, & = & \, S'_{z} \, + \, S_{xy} \, , \\
S'_{z}  \, & = & \, - \frac{ i J_{H} }{ 2 N } \, \sum_{k,p,q,\sigma} \,
\sigma \Biggl[ 
\frac{  \alpha^{\dagger}_{k \sigma} \alpha^{}_{p+q \sigma} 
\, \cos \eta_{k \sigma} \cos \eta_{p+q \sigma} }
{ E^{\alpha}_{k\sigma} - E^{\alpha}_{p+q \sigma} + \omega_q
 -  \omega_{k-p} } 
\, + \, 
\frac{ \beta^{\dagger}_{k \sigma} 
\beta^{}_{p+q \sigma} \, \sin\eta_{k\sigma} \sin\eta_{p+q \sigma}}
{E^{\beta}_{k\sigma} - E^{\beta}_{p+q \sigma} + \omega_q
 -  \omega_{k-p} } \,
 \nonumber \\
&  + & \,
\frac{ \alpha^{\dagger}_{k \sigma} 
\beta^{}_{p+q \sigma} \, \cos\eta_{k\sigma} \sin\eta_{p+q \sigma}}
{E^{\alpha}_{k\sigma} - E^{\beta}_{p+q \sigma} + \omega_q
 -  \omega_{k-p} } \,
 + \,
\frac{ \beta^{\dagger}_{k \sigma} 
\alpha^{}_{p+q \sigma} \, \sin\eta_{k\sigma} \cos\eta_{p+q \sigma}}
{E^{\beta}_{k\sigma} - E^{\alpha}_{p+q \sigma} + \omega_q
 -  \omega_{k-p} } \,  \,\Biggr] \,  b^{\dagger}_{q} b^{}_{k-p} \,, \\
S_{xy}  \, & = & \, - \frac{J_{H} \sqrt{M}}{i \sqrt{2 N}} \, 
\sum_{k,q} \,
\Biggl[ \frac{ \alpha^{\dagger}_{k \uparrow} 
\alpha^{}_{k+q \downarrow} \, \cos\eta_{k\uparrow} \cos\eta_{k+q \downarrow}}
{E^{\alpha}_{k\uparrow} - E^{\alpha}_{k+q \downarrow} + \omega_q} \, 
+ \, 
\frac{ \beta^{\dagger}_{k \uparrow} 
\beta^{}_{k+q \downarrow} \, \sin\eta_{k\uparrow} \sin\eta_{k+q \downarrow}}
{E^{\beta}_{k\uparrow} - E^{\beta}_{k+q \downarrow} + \omega_q} \,
\nonumber \\
& + & \,
\frac{ \alpha^{\dagger}_{k \uparrow} 
\beta^{}_{k+q \downarrow} \, \cos\eta_{k\uparrow} \sin\eta_{k+q \downarrow} }
{ E^{\alpha}_{k\uparrow} - E^{\beta}_{k+q \downarrow} + \omega_q  }\, 
+ \,
\frac{ \beta^{\dagger}_{k \uparrow} 
\alpha^{}_{k+q \downarrow} \, \sin\eta_{k\uparrow} \cos\eta_{k+q \downarrow}}
{E^{\beta}_{k\uparrow} - E^{\alpha}_{k+q \downarrow} + \omega_{q} } \, 
\Biggr]
\, b^{\dagger}_{q}  \,
\nonumber \\
& + & \, 
\Biggl[ \frac{ \alpha^{\dagger}_{k \downarrow} 
\alpha^{}_{k+q {\uparrow}} \, \cos\eta_{k\downarrow} \cos\eta_{k+q \uparrow}}
{E^{\alpha}_{k\downarrow} - E^{\alpha}_{k+q \uparrow} - \omega_{-q}} \, 
+ \, 
\frac{ \beta^{\dagger}_{k \downarrow} 
\beta^{}_{k+q \uparrow} \, \sin\eta_{k\downarrow} \sin\eta_{k+q \uparrow}}
{E^{\beta}_{k\downarrow} - E^{\beta}_{k+q \uparrow} - \omega_{-q}} \,
\nonumber \\
& + & \,
\frac{ \alpha^{\dagger}_{k \downarrow} 
\beta^{}_{k+q \uparrow} \, \cos\eta_{k \downarrow} \sin\eta_{k+q \uparrow}}
{E^{\alpha}_{k\downarrow} - E^{\beta}_{k+q \uparrow} - \omega_{-q} } \, 
+ \,
\frac{ \beta^{\dagger}_{k \downarrow} 
\alpha^{}_{k+q \uparrow} \, \sin\eta_{k\downarrow} \cos\eta_{k+q \uparrow}}
{E^{\beta}_{k\downarrow} - E^{\alpha}_{k+q \uparrow} - \omega_{-q} } \, 
\Biggr] \, b^{}_{-q}
\label{sresults}
\end{eqnarray}

Using Eq.~(\ref{h2}), through a rather lengthy calculation we have obtained the 
following second order effective spin-wave Hamiltonian, 
after projection onto the fermion ground state $\mid \Psi_F \rangle$:   
\begin{equation}
\tilde{H}^{(2)}_{sw} \, \equiv \, \langle \Psi_F \mid \tilde{H}^{(2)} \mid
\Psi_F \rangle \, = \, const. \, + \, \sum_{q} \, \tilde{\omega}_{q} \, 
b^{\dagger}_{q}  b^{}_{q} \, .
\end{equation}
It is interesting to notice that this Hamiltonian is directly  
diagonal in the original bare magnon operators, 
due to the fact that the projection onto states with fixed numbers of
$\alpha_{k \sigma}$ and $\beta_{k \sigma}$-electrons 
turns out to eliminate any non-diagonal contributions.    
The renormalization of the magnon energies, originating from the
coupling of the local Mn$^{4+}$ moments with the itinerant
electrons,  has contributions due to both the
longitudinal and the transverse parts of the Hund coupling
term. Specifically, we have obtained the following ``dressed'' spin
wave energies:
\begin{eqnarray}
\tilde{\omega}_{q} \, & = & \, \omega_{q} \, + \, \frac{J^{2}_{H} M}{2
  N}  \, \, t(q) \, \, + \, \,  \left( \frac{J_{H}}{2 N} \right)^{2} \,
  l(q) \, ,  
\label{wdressed}
\end{eqnarray}
where the transverse renormalization factor $t(q)$ could be viewed   
as describing virtual one-electron scattering processes  
with the creation or absorption of one magnon, depending on 
the detailed filling of the different bare electron spin subbands, and reads:
\begin{eqnarray}
t(q) \, & = & \, 
\sum_{k} \,
\Biggl[ \frac{ (n^{\alpha}_{k \uparrow} - n^{\alpha}_{k+q \downarrow})
 \, \cos^{2}\eta_{k\uparrow} \cos^{2}\eta_{k+q \downarrow}}
{E^{\alpha}_{k\uparrow} - E^{\alpha}_{k+q \downarrow} + \omega_q} \, 
+ \, 
\frac{ (n^{\beta}_{k \uparrow} - 
n^{\beta}_{k+q \downarrow}) \, \sin^{2}\eta_{k\uparrow} \sin^{2}\eta_{k+q \downarrow}}
{E^{\beta}_{k\uparrow} - E^{\beta}_{k+q \downarrow} + \omega_q} \,
\nonumber \\
& + & \,
\frac{ (n^{\alpha}_{k \uparrow} -  
n^{\beta}_{k+q \downarrow}) \, \cos^{2}\eta_{k\uparrow} \sin^{2}\eta_{k+q \downarrow} }
{ E^{\alpha}_{k\uparrow} - E^{\beta}_{k+q \downarrow} + \omega_q  }\, 
+ \,
\frac{ (n^{\beta}_{k \uparrow} 
-n^{\alpha}_{k+q \downarrow}) \, \sin^{2}\eta_{k\uparrow} \cos^{2}\eta_{k+q \downarrow}}
{E^{\beta}_{k\uparrow} - E^{\alpha}_{k+q \downarrow} + \omega_{q} } \, 
\Biggr] \, .
\label{tq}
\end{eqnarray}
The longitudinal magnon renormalization factor $l(q)$, on the other hand, 
involves virtual one electron-one magnon scattering processes, also depending  
on the detailed electron band fillings, and is given by:
\begin{eqnarray}
l(q) \, & = & \, \sum_{k,p,\sigma} \,
\sigma \Biggl[ 
\frac{ n^{\alpha}_{k \sigma} (1- n^{\alpha}_{p+q \sigma})  
\, \cos^{2} \eta_{k \sigma} \cos^{2} \eta_{p+q \sigma} }
{ E^{\alpha}_{k\sigma} - E^{\alpha}_{p+q \sigma} + \omega_q
 -  \omega_{k-p} } 
\, + \, 
\frac{ n^{\beta}_{k \sigma} (1 - n^{\beta}_{p+q \sigma} )\, 
\sin^{2}\eta_{k\sigma} \sin^{2}\eta_{p+q \sigma}}
{E^{\beta}_{k\sigma} - E^{\beta}_{p+q \sigma} + \omega_q
 -  \omega_{k-p} } \,
 \nonumber \\
&  + & \,
\frac{ n^{\alpha}_{k \sigma} (1-n^{\beta}_{p+q \sigma} )\, 
\cos^{2}\eta_{k\sigma} \sin^{2}\eta_{p+q \sigma}}
{E^{\alpha}_{k\sigma} - E^{\beta}_{p+q \sigma} + \omega_q
 -  \omega_{k-p} } \,
 + \,
\frac{ n^{\beta}_{k \sigma} (1-n^{\alpha}_{p+q \sigma}) \, 
\sin^{2}\eta_{k\sigma} \cos^{2}\eta_{p+q \sigma}}
{E^{\beta}_{k\sigma} - E^{\alpha}_{p+q \sigma} + \omega_q
 -  \omega_{k-p} } \,  \,\Biggr] \,  \,.
\label{lq}
\end{eqnarray}

\section{NUMERICAL RESULTS AND DISCUSSION}
\label{sec:results}

In this section we apply the general analytical results described above 
to the case of Tl$_{2}$Mn$_{2}$O$_{7}$, comparing the resulting spin excitations 
with the available experimental results.\cite{lynn} 

To evaluate the spin wave energies according to Eqs.~(\ref{wdressed})-(\ref{lq}), 
we need to perform sums over the first Brillouin zone (BZ) of the crystal lattice.
Here we have adopted a simplification, assuming the lattice to be a simple cubic one,    
as already done previously,\cite{civba,civmag} since the effect 
of the detailed lattice structure is not central to the present
discussion as we are studying ferromagnetic spin waves and considering a ferromagnetic 
superexchange interaction. This makes frustration effects due to the pyrochlore structure 
irrelevant.  

With this simplification, the bare magnon energies are given by:
\begin{equation}
 \omega_{q} \,  =  \, 2 J_{S} M \, [ \, ( 1 - \cos q_x a ) \, + 
\,  ( 1 - \cos q_y a ) \, +  ( 1 - \cos q_z a ) \, ] \, ,
\label{wbare}
\end{equation}
where $a$ is the lattice parameter.

To perform the BZ summations on the simple cubic lattice 
we have used the Chadi-Cohen process.\cite{ccohen,ccohen2} We noticed that to 
obtain dressed magnons with the appropriate symmetry of the lattice, it was not 
enough to use the basic wavevector set of the first BZ octant but that 
we needed to extend this set to the full first Brillouin zone, 
through the application 
of all the (48) symmetry operations of the $O_{h}$
group.\cite{tinkham}  
We already found good convergence using the third order of the Chadi-Cohen 
BZ summation process.\cite{ccohen,ccohen2}  

In Fig. 1 we show the bare isotropic ferromagnetic  magnon energies  along eight paths
 in the s.c. first BZ at temperature $T= 100 K$ 
, which would correspond to the results of the 
neutron scattering measurements.\cite{lynn} Regarding the dressed magnon energies, evaluated
 as described in the previous section, the general trend is that the interaction of
 the local moments  with the conduction  electrons causes a softening of the ferromagnetic
 spin waves. The amplitude of the effect is however very 
small, consistently with the experimental data.\cite{lynn} Of the two contributions 
to the renormalization, the longitudinal one (Eq.~(\ref{lq})) is almost negligible, 
being  at least four orders of magnitude smaller than the bare energies.
 Therefore the renormalization seen in Fig. 1 is essentially due 
 to the transverse contributions (Eq.~(\ref{tq})). 
It is worth mentioning that in Fig. 1 we are extending our  
perturbation calculation well beyond its expected range of convergence,
 because we choose $ \mid J_H/J_S \mid > 1 $ as was found appropriate
 for Tl$_{2}$Mn$_{2}$O$_7$.\cite{civmag} 
 This causes problems near $q=0$, where the bare magnon energies are smaller
 than their negative renormalization contribution. 
Away from $q=0$, we find a remarkable overall agreement 
in the whole BZ between the  dressed ferromagnetic magnons which we obtain 
 and the bare spin waves. 
This we ascribe to the fact that there are only few conduction electrons present 
in this compound,\cite{5,7,civba,civmag} 
 as can be also seen in Fig. 2 depicting the band energies and 
 indicating the Fermi level.  
There we see that the up-spin subbands for both $\alpha$ and $\beta$ electrons  are 
always higher in energy than their spin-down counterparts as expected
 from Eq. (8) since 
we have used  $J_H < 0$. Also, the $\beta$ states are almost empty 
 while the down-spin $\alpha$ band is almost completely full.  

It is interesting to mention that analyzing the effect of temperature on the dressed 
magnons calculated, we verified that we obtain the same trend which has been observed 
in the experiments,\cite{lynn} namely, we do get a softening of the spin excitations 
as temperature increases though quantitatively our temperature effect  
is about one order of magnitude smaller than the measured one. 

The following five figures have been included to allow a better  understanding of the 
influence of the different parameters on the spin wave renormalization. We plot the results 
for a subset of non-equivalent wavevectors to allow visualization of the  $q$-dependence 
of the renormalization.  
 
Figs. 3 and 4 show the variation of the renormalized spin wave energy 
relative to its bare value versus, respectively, $J_S$ and $J_H$, for a
 subset of seven non-equivalent wavevectors selected in the BZ. 
These two figures show opposite trends, which can be easily understood. As the magnitude of $J_H$ 
grows with respect to that of $J_S$, the local moment system interacts more strongly with
 the itinerant electrons and the corresponding magnon energy acquires a larger 
renormalization. 
Conversely, if $J_S$ grows with respect to $J_H$, the electronic effect on the 
magnons decreases, because the exchange energy now has the main role in determining the 
spin wave energy. 

The following two figures show the effects of varying the
 electronic band parameters. While Fig. 5 shows that the renormalized magnon energy is rather
 insensitive to the hybridization $t_{cd}$ (except for one of the wavevectors plotted), 
in Fig. 6 we observe that the bandwidth ratio $t_d/t_c$ 
can have a sizeable effect, at least for wavectors away from the BZ boundaries.
This effect might perhaps be explored experimentally in neutron scattering measurements 
under pressure. 

In Fig. 7 we exhibit the effect of electron filling (even quite far above the 
ranges appropriate for the description of   Tl$_{2}$Mn$_{2}$O$_7$ \cite{civba,civmag}).
As one would expect, the dressed magnon energies are quite sensitive to the
electron filling, with quite larger magnitudes of spin wave softenings,
monotonously increasing with $n$ in the range showed.    

Two  common features are present in Figs. 3-7: one, that the renormalization generally 
results in a softening of the spin excitations by their coupling with the electrons; 
the other, that its magnitude is larger for wavevectors away 
from the Brillouin zone boundaries.

Our last figure (Fig. 8), was included to exhibit the effect of a change of sign of 
the Hund coupling $J_{H}$ on the renormalized magnons, in this case favouring FM 
alignment of the d-electron spins with the local spins (it is worth mentioning that 
the ground state of the generic model is symmetric with respect to such sign change 
if accompanied by a reversal of the relative spin orientations of the local moments
and $d-$spins, as discussed in Ref. 18). Such sign change 
reverses the relative order of the spin up and down electron eigenbands: 
the longitudinal magnon renormalization contribution (Eq. (22)) therefore changes sign; 
the transversal contribution (see Eq. (21)), dominating the magnon renormalization 
as mentioned above, changes in a less obvious way under sign reversal of the Hund coupling,
leading to the dressed magnons depicted in Fig. 8.  Here a hardening of the magnons  
away from the long wavelenghts' region is visible, contrary to the softening observed 
on the whole BZ when the Hund coupling favours antiparallel alignment 
of the d-electron spins with respect to the local spins. This is an interesting prediction, 
being the discussion of the sign of such Hund coupling in  Tl$_2$Mn$_2$O$_7$
 a non-settled matter (e.g. in Refs. 21 and 6 an antiparallel alignment due to the 
Hund coupling is mentioned, while e.g. in Refs. 16 and 26 the opposite case is proposed),
as it would provide a tool to determine experimentally such sign 
by measuring the magnons with neutron scattering experiments on Tl$_{2}$Mn$_{2}$O$_7$ 
performed in the yet unexplored shorter wavelenghts' region of the BZ.

\section{SUMMARY}
\label{sec:summary}

In the present work, we have addressed the problem of the spin excitations 
measured in Tl$_2$Mn$_2$O$_7$,\cite{lynn} found to exhibit features characteristic of 
isotropic ferromagnets and contrasting starkly with the results in CMR perovskites.  
Using the generic model \cite{civmag} to describe the compound, assuming a ferromagnetic 
superexchange interaction between the Mn$^{4+}$ spins,  
we have employed a  perturbative approach to calculate 
the magnon dispersion relation along different paths in the first Brillouin zone. 
In such way, we were able to describe qualitatively the available experimental results,
using parameters for the generic model as found appropriate for the description 
of Tl$_2$Mn$_2$O$_7$ before.\cite{civmag}  

Even though in the present work we are extending our
perturbative calculation well beyond its expected range of
convergence by the magnetic coupling parameters used (resulting
in problems near $q=0$: where bare magnon energies are smaller than
their renormalization), a remarkable overall agreement along the
Brillouin zone is evident between the dressed ferromagnetic magnons 
obtained from the generic model and the bare spin waves
(characteristic of an ideal isotropic ferromagnet,   
such as found in experiments.\cite{lynn}) This is due to the
electronic structure particularities and the small number of effective carriers
 in Tl$_2$Mn$_2$O$_7$.\cite{7,civba,sushkolast,civmag} A measurement of the magnon 
dispersion relations at smaller wavelenghts, should allow to distinguish  
if the Hund coupling favours parallel or antiparallel alignment of the 
Mn$^{4+}$ spins and the narrow band electrons, according to the present predictions
of a hardening or softening, respectively, of the dressed magnons in such cases.    
According to our results, there should also be a sizeable effect of pressure 
upon the spin excitations, corresponding to the ensuing variation of 
the relative electron bandwidths of the generic model, which it would also
be interesting to study experimentally.

\acknowledgments 

C.I.V. thanks Jeff Lynn for drawing her attention to this problem  
and for enlightening discussions. Both authors thank B.Alascio for 
his interest and comments. C.I.V. is a member of the Carrera del
Investigador Cient\'{\i}fico of CONICET (Consejo Nacional de
Investigaciones Cient\'{\i}ficas y T\'ecnicas, Argentina), and wishes to 
acknowledge the financial support and hospitality of the 
Dipartimento di Fisica, Universit\'a di Parma, and International
Centre for Theoretical Physics, Trieste, 
where part of this work was done. M.A. acknowledges financial 
support of I.N.F.M. and the hospitality of Centro At\'omico Bariloche.

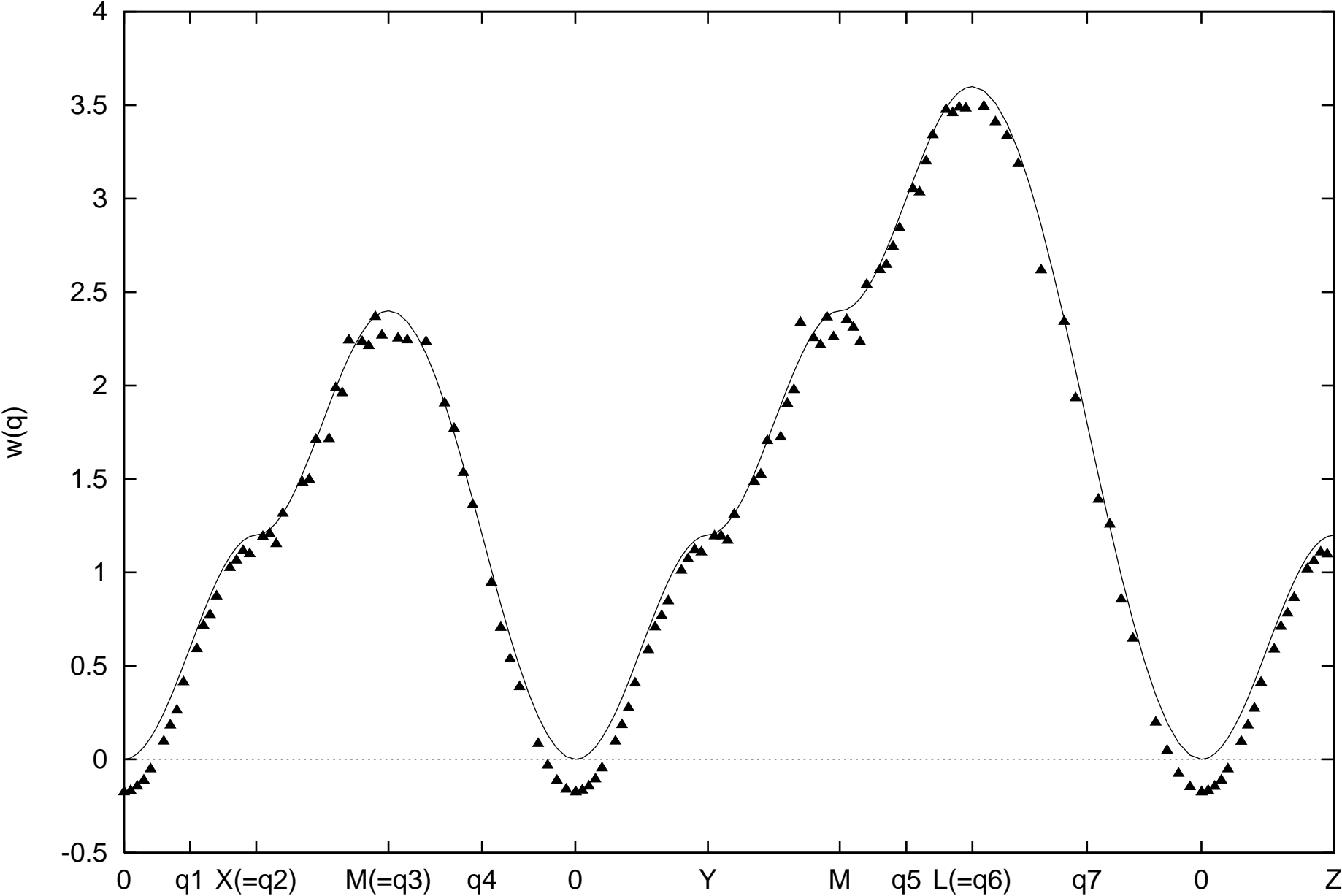
\begin{figure}
\caption{ Spin wave dispersion relations along eight trajectories in
the s.c. first Brillouin zone: bare ferromagnetic magnons (solid line) 
vs. dressed magnons calculated (points).   
Energies in units of $t_c$; origin: $\epsilon_c = 0 $.
Other parameters of the generic model for Tl$_2$Mn$_2$O$_7$:  
$\epsilon_d = - 0.7 $; $t_{d} = 0.1 $;    
$ t_{cd} = 0.3 $; $ n = 1.07 $; 
$J_{S} = 0.2$ ; $ J_{H} =  -0.9$ .
BZ points: $ 0= \Gamma = (0,0,0)$,  $X=(\pi/a,0,0)$, $Y=(0,\pi/a,0)$,
 $Z=(0,0,\pi/a)$, $M=(\pi/a,\pi/a,0)$ and 
$L=(\pi/a,\pi/a,\pi/a)$.}
\label{fig:magnons}
\end{figure}
    
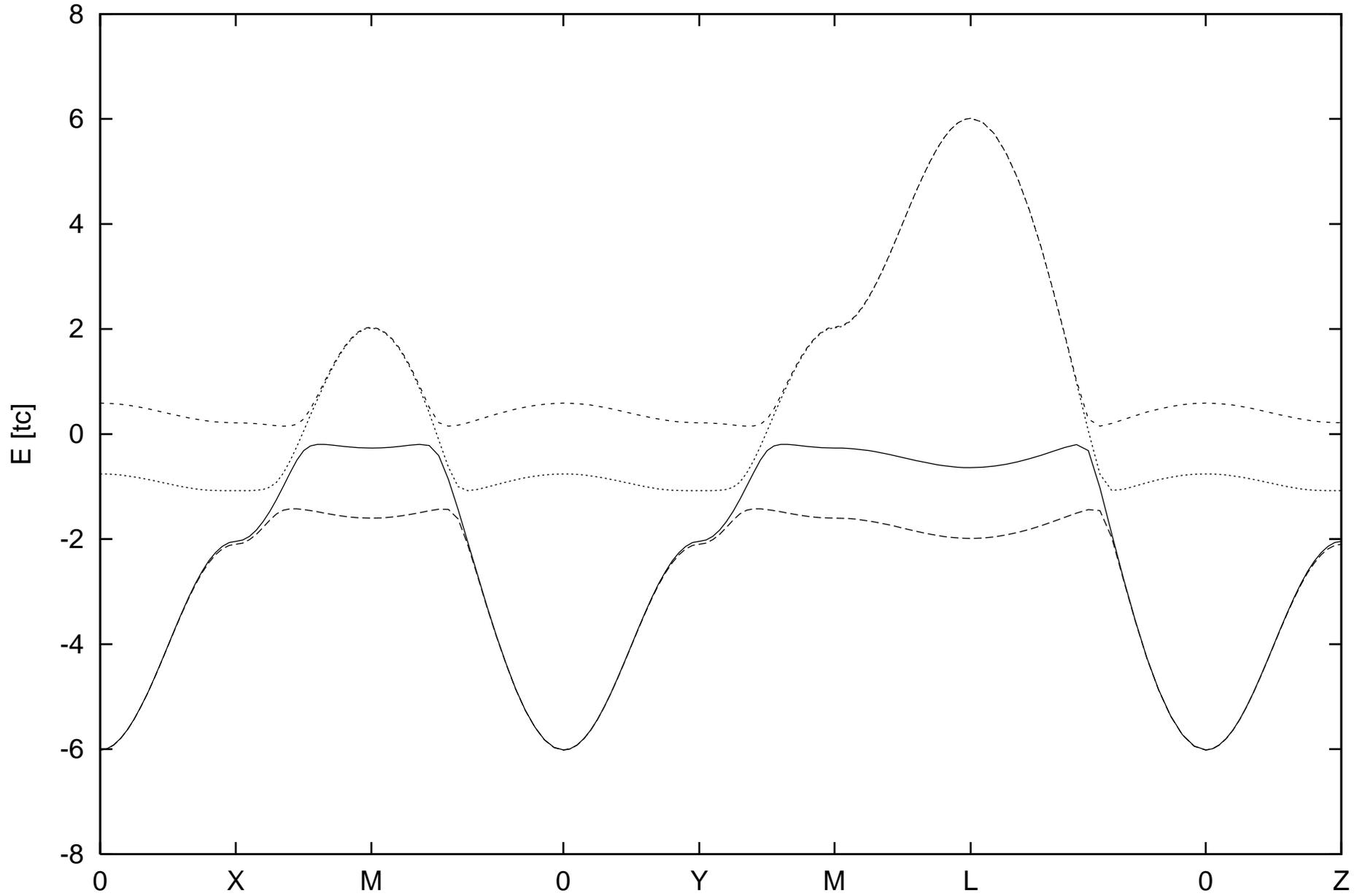
\begin{figure}
\caption{ Bare electronic spin subband dispersion relations, 
along eight trajectories in the s.c. first Brillouin zone: 
$E_{\alpha \uparrow}$ (solid); $E_{\alpha \downarrow}$ (long dashes);  
$E_{\beta \uparrow}$ (short dashes); $E_{\beta \downarrow}$ (dotted). 
Fermi level: $ \mu / t_c   = -1.47 $. 
BZ points and parameters as in Fig.~\ref{fig:magnons}.}    
\label{fig:bands}
\end{figure}

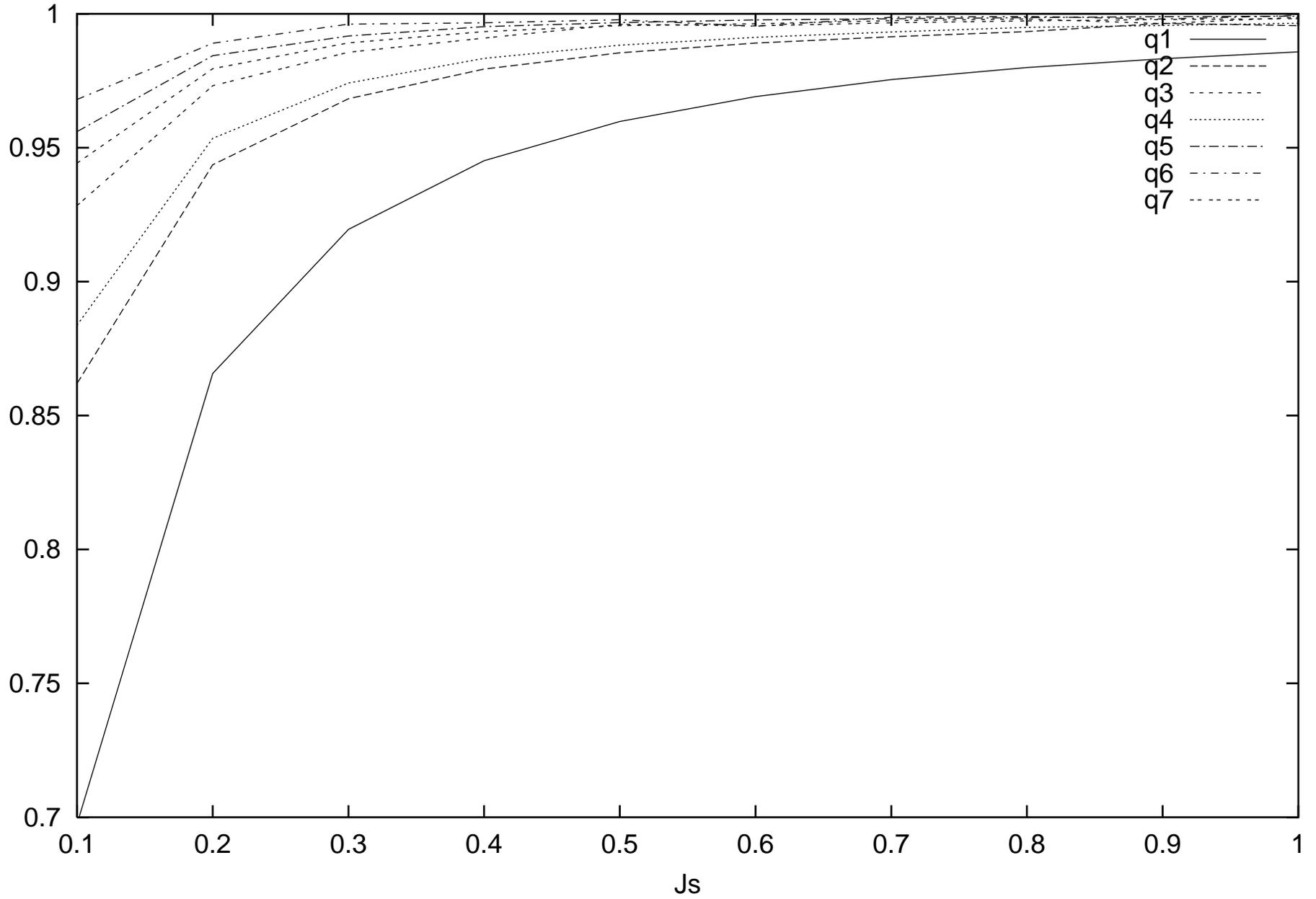
\begin{figure}
\caption{ Magnon renormalization factor ($= \tilde{\omega}_{q} / \omega_{q}$)  
vs. superexchange coupling, $J_{S}$,  
for seven non-equivalent wave-vectors 
(indicated as $q_{1} - q_{7}$ along the abcissa of Fig.~\ref{fig:magnons}).
$ J_{H} = -0.8 $; other parameters as in Fig.~\ref{fig:magnons}.}
\label{fig:jsvar}
\end{figure}

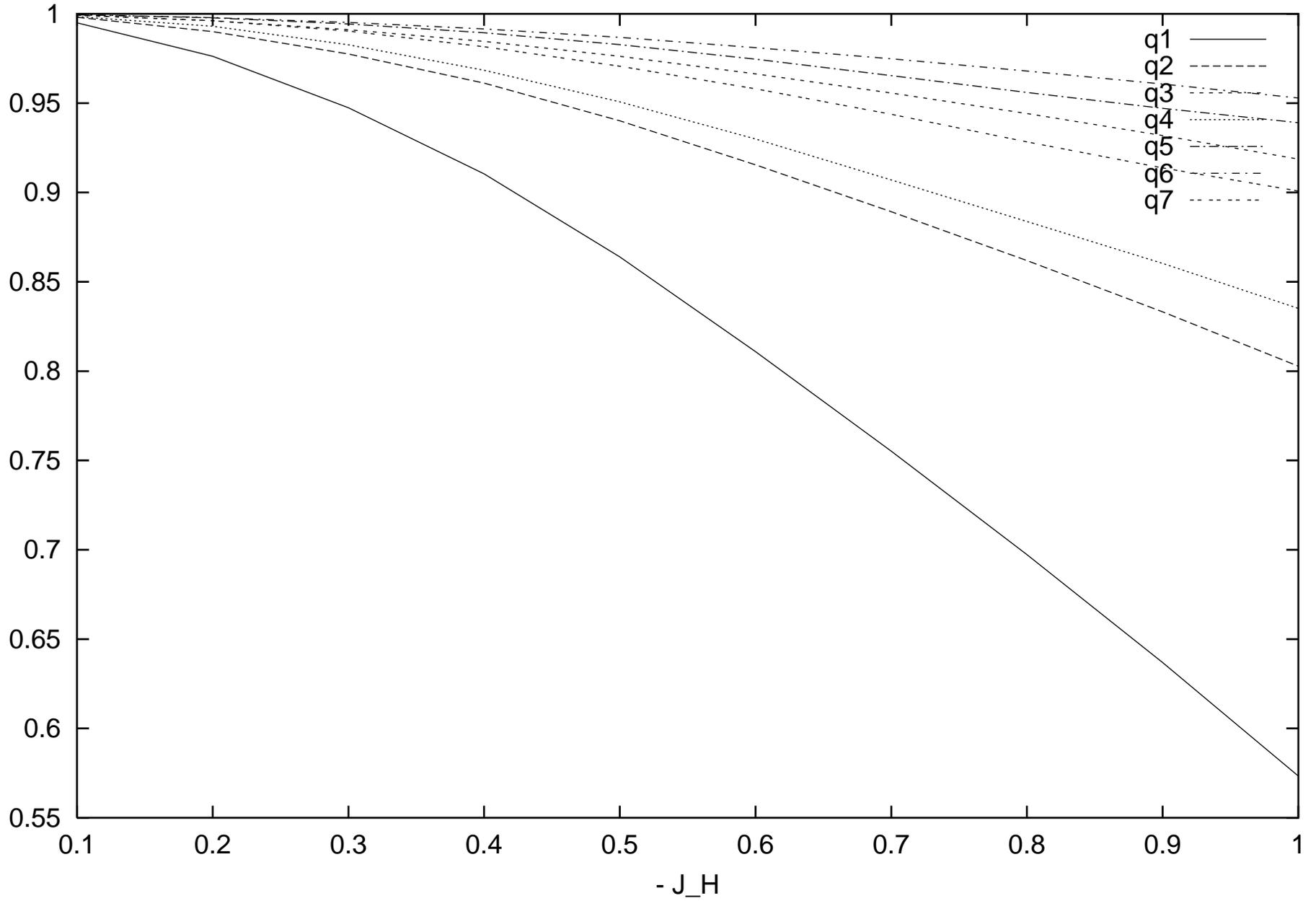
\begin{figure}
\caption{ $\tilde{\omega}_{q} / \omega_{q}$ vs. magnitude 
 of the Hund coupling, $\mid J_{H} \mid$, 
for the same non-equivalent wave-vectors of Fig.~\ref{fig:jsvar}. 
$ J_{S} = 0.1 $; other parameters as in Fig.~\ref{fig:magnons}.}
\label{fig:jhvar}
\end{figure}

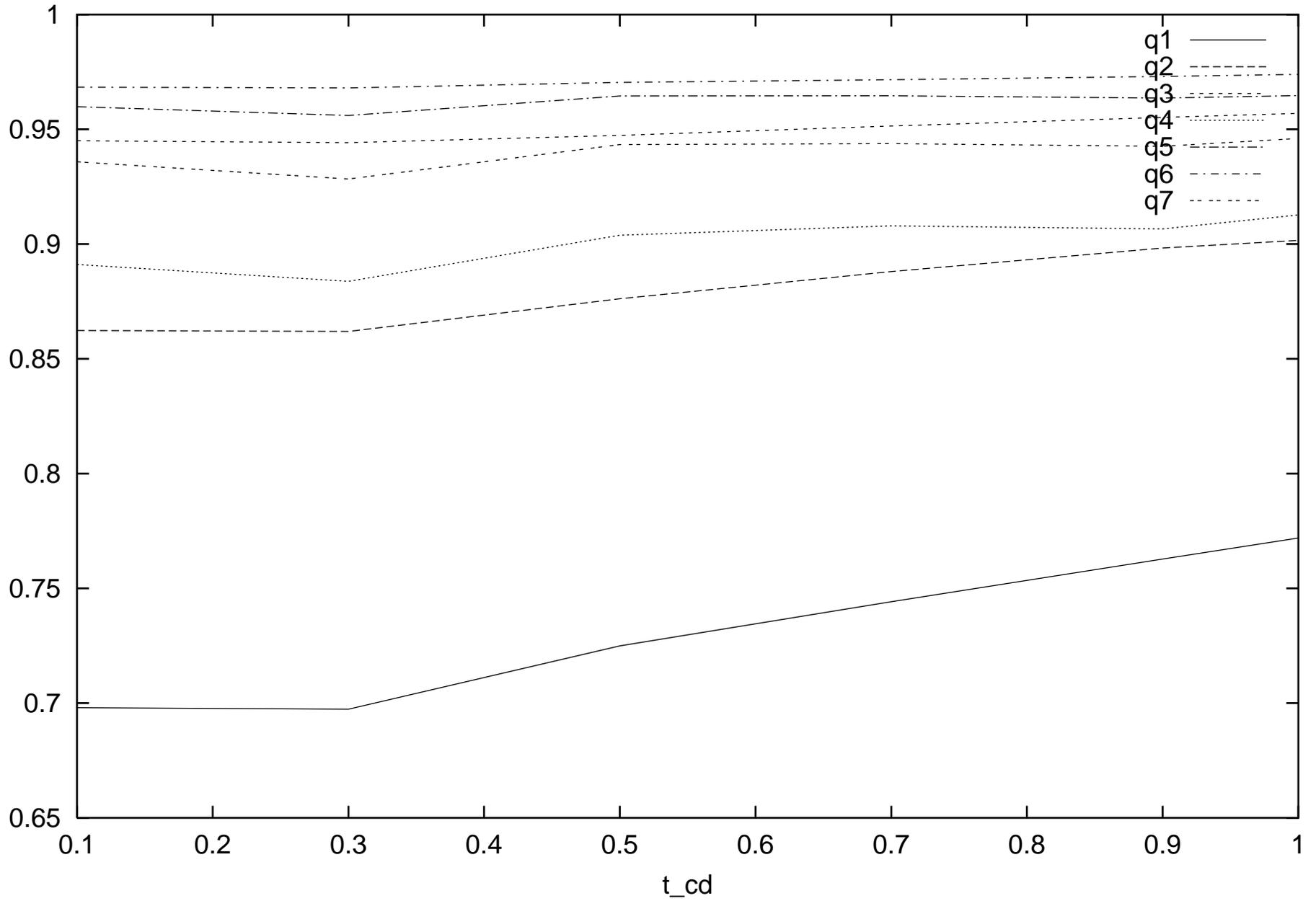
\begin{figure}
\caption{ $\tilde{\omega}_{q} / \omega_{q}$ vs.  hybridization,
$t_{cd}$, for the same non-equivalent wave-vectors 
of Fig.~\ref{fig:jsvar}. $ J_{S} = 0.1; J_{H} = -0.8$; 
other parameters as in Fig.~\ref{fig:magnons}.}
\label{fig:vvar}
\end{figure}

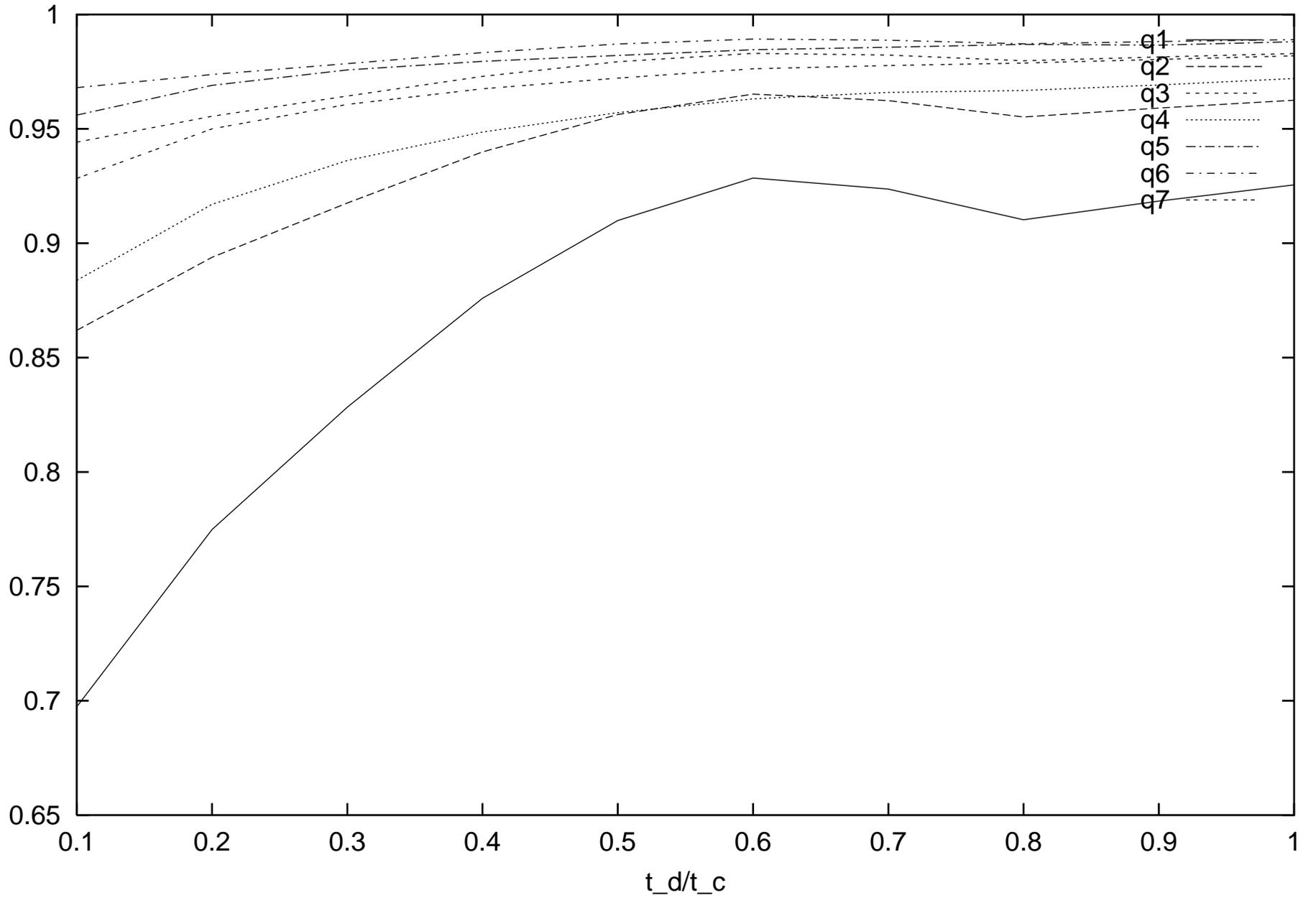
\begin{figure}
\caption{ $\tilde{\omega}_{q} / \omega_{q}$ 
vs. relative bandwidth, $t_{d}/t_{c}$, 
for the same non-equivalent wave-vectors 
of Fig.~\ref{fig:jsvar}. Other parameters as in Fig.~\ref{fig:vvar}.}
\label{fig:alphavar}
\end{figure}

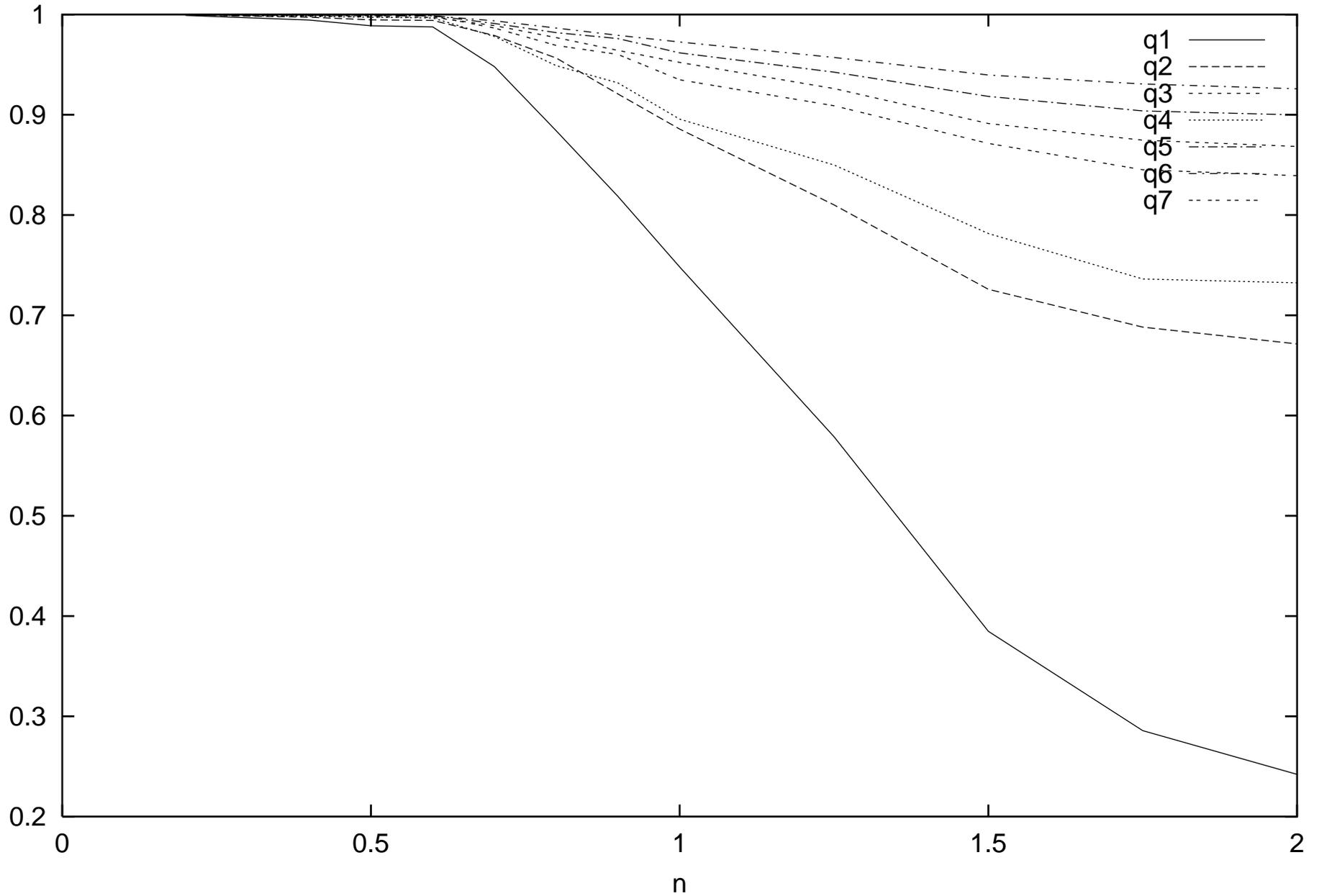
\begin{figure}
\caption{ $\tilde{\omega}_{q} / \omega_{q}$  
 vs. total electron filling, $n$, 
for the same non-equivalent wave-vectors 
of Fig.~\ref{fig:jsvar}. Other parameters as in Fig.~\ref{fig:vvar}.}
\label{fig:nvar}
\end{figure}

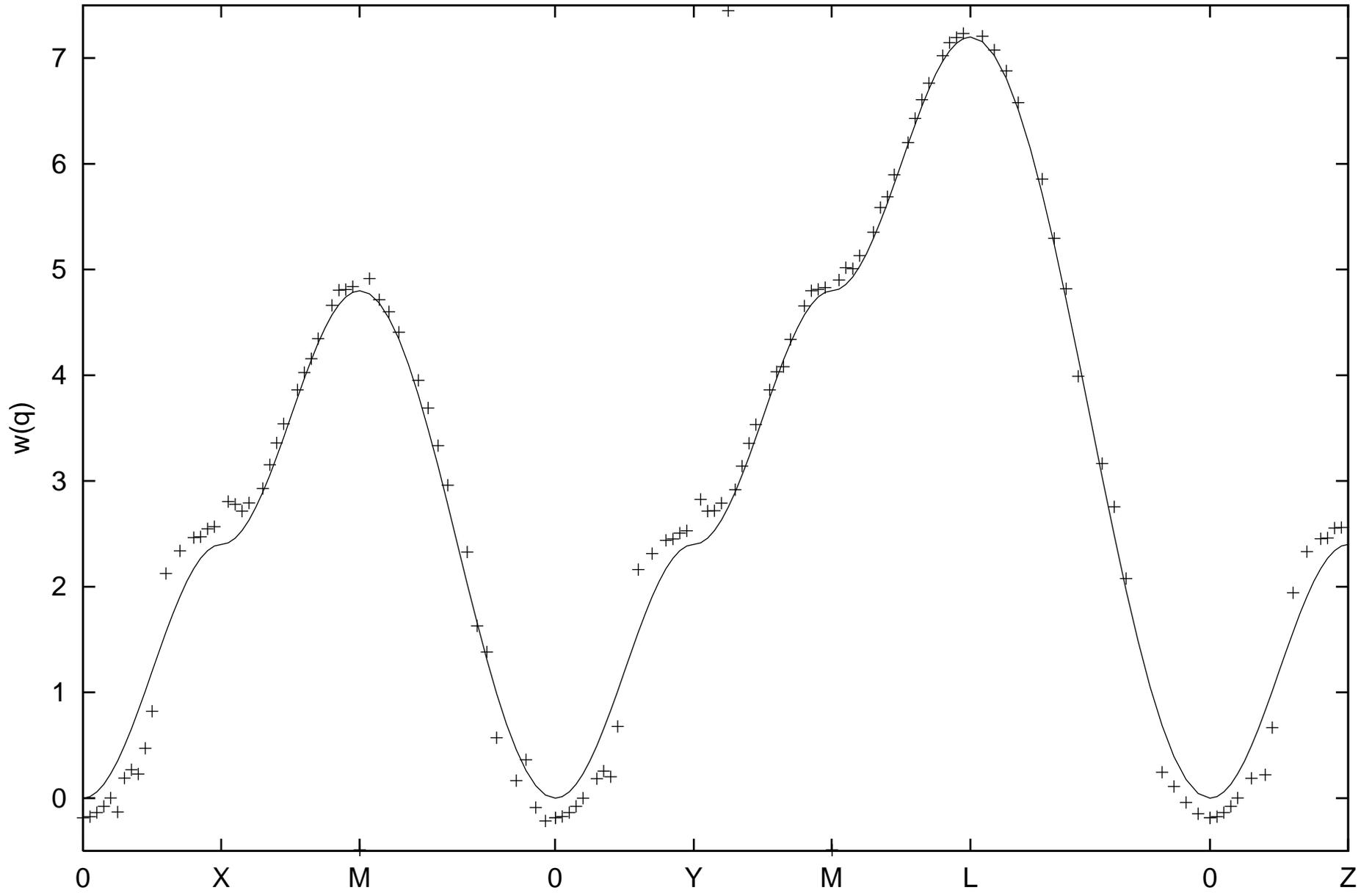
\begin{figure}
\caption{ Spin wave dispersion relations along eight trajectories in
the s.c. first BZ: bare ferromagnetic magnons (solid line) 
vs. dressed magnons calculated (points).  $ J_{H} =  +0.8$; $J_{S} = 0.4$;   
other parameters as in Fig.~\ref{fig:magnons}.
}
\label{fig:nmagnons}
\end{figure}


\begin{references}

\bibitem{5} Y. Shimakawa, Y. Kubo, and T. Manako, Nature 379, 55(1996);
Y. Shimakawa, Y. Kubo, T. Manako, Y. V. Sushko, D. N. Argyriou, and 
J. D. Jorgensen, Phys. Rev. B {\bf 55}, 6399 (1997).  
\bibitem{6} S. W.Cheong, H. Y. Hwang, B. Batlogg, and L. W. Rupp Jr., 
Solid State Commun. {\bf 98}, 163 (1996).
\bibitem{7} M. A. Subramanian, B. H. Toby, A. P. Ramirez, W. J. Marshall,  
A. W. Sleight, and G. H. Kwei, Science {\bf 273}, 81 (1996).
\bibitem{3} R. von Helmolt, J. Wecker, B. Holzapfel, L. Schultz, and 
K. Samwer, Phys. Rev. Lett. {\bf 71}, 2331 (1993); S. Jin, T. H. Tiefel,
M. McCormack, R. A. Fastnacht, R. Ramesh, and L. H. Chen, Science {\bf
264}, 413 (1994).
\bibitem{single} G. H. Kwei, C. H. Booth, F. Bridges, and
M. A. Subramanian, Phys. Rev. B {\bf 55}, R688 (1997).
\bibitem{sushkolast} H. Imai, Y. Shimakawa, Y. V. Sushko, and Y. Kubo, 
Phys. Rev. B {\bf 62}, 12190 (2000). 
\bibitem{presion} Y. V. Sushko, Y. Kubo, Y. Shimakawa, and T. Manako,
Czech. J. Phys. {\bf 46}, 2003 (1996); Y. V. Sushko, Y. Kubo,
Y. Shimakawa, and T. Manako, Rev. High Pressure Sci. Tech. {\bf 7},
505 (1998); Y. V. Sushko, Y. Kubo, Y. Shimakawa, and T. Manako,
Physica B {\bf 259-261}, 831 (1999).
\bibitem{boston} M. N\'u\~nez Regueiro {\em et al.}, in {\it
Proceedings of the MRS Meeting, Boston, 1999}.
\bibitem{rudoping} T. Takeda {\em et al.}, J. Sol. State Chem. {\bf
140}, 182 (1998); B. Mart\'{\i}nez {\em et al.}, Phys. Rev. Lett. {\bf 83},
2022 (1999); R. Senis {\em et al.}, Phys. Rev. B {\bf 61}, 11637
(2000).
\bibitem{scdoping} A. P. Ramirez and M. A. Subramanian, Science {\bf
277}, 546 (1997).
\bibitem{sbdoping} J. A. Alonso {\em et al.}, Phys. Rev. B {\bf 60},
15024 (1999).
\bibitem{bidoping} J. A. Alonso, J. L. Mart\'{\i}nez,
M. J. Mart\'{\i}nez-Lope, M. T. Casais, and
M. T. Fern\'andez-D\'{\i}az, Phys. Rev. Lett. {\bf 82}, 189 (1999).
\bibitem{Odefect} J. A. Alonso, M. J. Mart\'{\i}nez-Lope, M. T. Casais,
J. L. Mart\'{\i}nez, and M. T. Fern\'andez-D\'{\i}az, 
Chemistry of Materials {\bf 12}, 1127 (2000).
\bibitem{cddoping} J. A. Alonso {\em et al.}, Appl. Phys. Lett. {\bf
76}, 3274 (2000); P. Velasco, J. A. Alonso,  M. T. Casais, 
M. J. Mart\'{\i}nez-Lope, J. L. Mart\'{\i}nez, and
M. T. Fern\'andez-D\'{\i}az, Phys. Rev. B {\bf 66}, 174408 (2002). 
\bibitem{civba} C. I. Ventura and B. Alascio, Phys. Rev. B {\bf
56}, 14533 (1997); \, C. I. Ventura and B. Alascio, in {\it Current Problems in
Condensed Matter}, edited by J. L. Mor\'an-L\'opez (Plenum Press, 
New York, 1998), p. 27.
\bibitem{littlewood} P. Majumdar and P. B. Littlewood,
Phys. Rev. Lett. {\bf 81}, 1314 (1998).
\bibitem{dolores} M. D. N\'u\~nez-Regueiro and C. Lacroix,
Phys. Rev. B {\bf 63}, 14417 (2001).
\bibitem{civmag} C.I.Ventura and M.A.Gusm$\tilde{a}$o,  
Phys. Rev. B {\bf 65}, 14422 (2002).
\bibitem{bsub} D. K. Seo, M. H. Whangbo, and M. A. Subramanian, 
Solid State Commun. {\bf 101}, 417 (1997).  
\bibitem{bsingh} D. J. Singh, Phys. Rev. B {\bf 55}, 313 (1997).
\bibitem{mishra} S. K. Mishra and S. Satpathy, Phys. Rev. B {\bf 58},
7585 (1998).
\bibitem{lynn} J.W.Lynn, L.Vasiliu-Doloc and M.A. Subramanian, 
Phys. Rev. Lett. {\bf 80},4582(1998).
\bibitem{ccohen} D.J. Chadi and M.L.Cohen, Phys. Rev. B {\bf 8}, 5747 (1973). 
\bibitem{ccohen2} L.Macot and B.Frank,  Phys. Rev. B {\bf 41}, 4469 (1990).
\bibitem{tinkham} M.Tinkham, ``Group Theory and Quantum Mechanics'', 
McGraw-Hill Book Co., New York (1964).
\bibitem{guinea}P. Velasco {\em et al.}, Phys. Rev. B {\bf 66}, 104412
 (2002). 


\end{references}
\end{document}